\documentclass[aps,preprint]{revtex4-1}
\usepackage{graphicx}
\usepackage{epsfig}
\begin{document}
\title{Superradiance and black hole bomb in five-dimensional minimal ungauged supergravity}
\author{\large Alikram N. Aliev}
\address{Faculty of Engineering and Architecture, Yeni Y\"{u}zy{\i}l University, Cevizliba\v{g}-Topkap{\i}, 34010  Istanbul, Turkey}
\date{\today}

\begin{abstract}

We examine the black hole bomb model which consists of a rotating black hole of five-dimenensional minimal  ungauged supergravity and a reflecting mirror around it. For low-frequency scalar perturbations, we find  solutions to the Klein-Gordon equation in the near-horizon and far regions of the black hole spacetime. To avoid  solutions with logarithmic terms, we assume that the orbital quantum number  $ l  $ takes on nearly, but not exactly, integer values and perform the matching of these solutions in an intermediate region. This allows us to calculate analytically the frequency spectrum of quasinormal modes,  taking  the limits as $ l  $  approaches even or odd integers separately. We find that all $ l  $ modes of scalar perturbations undergo negative damping in the regime of superradiance, resulting in exponential growth of their amplitudes. Thus, the model under consideration would exhibit the superradiant instability, eventually behaving  as a black hole bomb in five dimensions.

\end{abstract}


\maketitle

\section{Introduction}

The phenomenon of {\it superradiance} through which waves of certain frequencies  are amplified when  interacting with a medium has long been known in both classical and quantum non-gravitational systems. The quantum aspect of this phenomenon traces back to the so-called Klein paradox \cite{klein, sauter} whose  subsequent resolution revealed the existence of superradiant boson (not fermion) states in the background of strong electromagnetic fields (see e.g. \cite{mano} and references therein).  The superradiant effect also arises in many  classical systems moving  through a medium with the linear  velocity that exceeds the  phase velocity of waves under consideration.  As early as 1934 it was known that the reflection of sound waves from  the boundary of a medium, which moves with supersonic velocity, occurs with amplification \cite{ar}. Subsequently, examples of such an amplification were found  in a number of cases; for instance, in the motion of carriers in an elastic piezoelectric  substance \cite{white}  as well as in the motion of a conducting liquid in a resonator \cite{andron}.

Zel'dovich  first  realized that the superradiant  condition can be fulfilled  in a rotational case as well \cite{zeldovich}.  Suggesting that for a wave of frequency $\omega $ and angular momentum $ m $, the angular velocity $\Omega $
of a body can exceed the angular phase velocity $\omega/m $ of the wave,  $\Omega  > \omega/m $, he demonstrated  the amplification of waves reflected from a rotating and conducting cylinder. In addition, Zel'dovich put forward the idea that a semitransparent mirror surrounding the cylinder could provide  exponential amplification of waves. He  also anticipated that the phenomenon of superradiance  and the process of  exponential amplification of  waves would occur in the field of a Kerr black hole.  The black  hole superradiance  was independently predicted by Misner \cite{misner}, who pointed out that certain modes of scalar waves scattered off the  Kerr  black hole undergo amplification. Possible applications of the superradiant mechanism were explored by  Press and Teukolsky \cite{press1}. In particular, by locating a spherical mirror around a rotating black hole they pointed out that  such a system  would eventually develop a strong instability against exponentially growing modes in the superradiant regime, thus creating  a {\it black hole bomb}.

The quantitative theory of superradiance for scalar, electromagnetic and gravitational  waves in the Kerr metric  was developed  in classic papers  by  Starobinsky \cite{starobinsky} and Starobinsky  and Churilov \cite{starchur}  (see also \cite{press2, press3}). The existence of superradiance is intimately related to the salient feature of the Kerr metric; the timelike Killing vector that defines the energy with respect to asymptotic  observers becomes spacelike in the region located outside the horizon, called the {\it ergoregion}. This in turn entails the possibility of  negative energy states within the ergoregion, underpinning  the physical interpretation  of the superradiant effect. Scattering a wave off a rotating black hole may cause fluctuations of the negative energy states, resulting in the negative energy flux into the black hole \cite{unruh1}. As a consequence,  the scattered wave  becomes amplified, by conservation of energy. It should be noted that there is no superradiance for fermion modes in the Kerr metric, as  shown by detailed calculations in \cite{unruh2, rahmi}.

The black hole superradiance on its own  has only a conceptual significance,  showing the possibility of the extraction of rotational energy from the black hole due to the wave mechanism. However, it has played a profound role in addressing  the stability issues of rotating black holes in general relativity, against  small external perturbations. Developments in this direction have revealed  that rotating black holes are stable to massless scalar, electromagnetic and gravitational perturbations \cite{press2, press3}. On the contrary, it appeared that small perturbations of a massive scalar field  grow exponentially in the superradiant  regime,  creating  the instability of the system,  the black hole bomb effect \cite{damour, detweiler, dolan}. The physical reason underlying  this effect is that the  motion of a massive particle around a rotating black hole may  occur in stable circular orbits \cite{bardeen} (see also a recent paper \cite{ag}). Thus, to view the instability one can imagine a wave-packet of the massive scalar field moving in these orbits and forming  ``bound states" in the  well of the effective potential of the motion. Though the potential barrier keeps the wave-packet bound states in the well from escaping to infinity, but  from quantum-mechanical point of view they would tunnel through the barrier into the horizon. As a consequence, the  bound states in the well become {\it quasinormal} with  complex frequencies whose imaginary parts in the superradiant regime determine the growth rate of the  wave-packet modes.  It is clear that the runaway behavior of such modes between the potential well and the horizon would result in their  continuous  reamplification, thereby causing the instability.

Another realization of the black hole bomb effect occurs in anti-de Sitter (AdS) spacetimes. This is due to the fact that in the regime of superradiance, the timelike boundary of the AdS spacetime plays the role of a resonant cavity between the black hole and spatial infinity. In \cite{hreall}, it was argued that small rotating AdS black holes in five dimensions may exhibit the superradiant instability against  external perturbations. The authors of works \cite{cardoso1, cardoso2} were the first to develop these arguments further by using both analytical and numerical approaches. Elaborating on the black hole bomb effect of Press and Teukolsky in four dimensions,  they pointed out that its realization  crucially depends on the distance at which  the mirror must be located. Thus,  for the superradiant modes to be excited there  exists a critical radius and below this radius the system  is stable. These results  allow one to  clarify the  instability of small Kerr-AdS black holes, as discussed  in \cite{cardoso2}. Continuing this line of investigation in five  dimensions, the superradiant instability of small rotating charged AdS black holes was considered in \cite{ad}. Meanwhile, the case of arbitrarily higher dimensions for  small Reissner-Nordstr\"{o}m-AdS  black holes has recently been studied in \cite{hwang}. In particular, it was noted that for some values of the orbital quantum number $ l $, which can occur in  odd spacetime  dimensions,  the  analytical approach  of \cite{ad} fails being responsible for the seeming absence of the superradiant instability for certain modes. Detailed numerical  calculations have  shown  that the superradiant  instability exists in all higher dimensions and with respect to all modes of scalar perturbations \cite{hwang}.

In this paper, we wish to embark on a further exploration of the superradiant instability for rotating black holes in five dimensions. We consider the black hole bomb model for scalar perturbations, which consists of a rotating black hole of five-dimenensional minimal ungauged supergravity and a reflecting mirror around it. In Sec. II we begin by discussing the defining properties of the  spacetime metric for the black hole under consideration. Here we present remarkably simple formulas for the coordinate angular velocities of locally nonrotating observers.  These formulas reveal the ``bi-dragging" property of the black hole at large distances and reduce to its angular velocities as one approaches the horizon. Next, we introduce a corotating Killing vector field  which is tangent to the null geodesics of the horizon and calculate the surface gravity and the electrostatic potential of the horizon. In Sec. III we discuss the separated  radial and angular parts of the Klein-Gordon equation for  a charged  massless scalar field and derive the threshold inequality for superradiance. Focusing on low-frequency perturbations, in Sec. IV  we  find  solutions to the radial wave equation by dividing the spacetime into the near-horizon and far regions. To avoid solutions with logarithmic terms, we then assume that the orbital quantum number $ l  $  is an approximate integer and perform the matching of these solutions in an intermediate region. In Sec. V  we  calculate the frequency spectrum of quasinormal modes in the black hole-mirror system, taking  the limits as $ l $  approaches even or odd integers separately. Here we show that in the regime of superradiance,  the black hole-mirror system exhibits instability to all $ l $ modes of scalar perturbations. In Sec. VI  we end up with a discussion of our results.

\begin{center}
\section{The metric}
\end{center}

The general solution to five-dimensional minimal gauged supergravity that describes charged and rotating black holes with two independent rotational symmetries was found  by Chong, Cveti\u{c}, L\"{u} and Pope (CCLP) \cite{cclp}. In the case of ungauged supergravity (the vanishing cosmological constant) it is given by the metric
\begin{eqnarray}
\label{gsugrabh}
ds^2 & = & - \left( dt - a
\sin^2\theta\,d\phi - b
\cos^2\theta\,d\psi \right)\nonumber
\left[f\left( dt - a
\sin^2\theta\,d\phi\ - b
\cos^2\theta\,d\psi \right)\nonumber
\right. \\  & & \left. \nonumber
+ \frac{2 Q}{\Sigma}\left(b
\sin^2\theta\,d\phi + a
\cos^2\theta\,d\psi \right)\right]
+ \Sigma\left(\frac{r^2 dr^2}{\Delta} + d\theta^{\,2}\right) \\ &&
+ \frac{\sin^2\theta}{\Sigma} \left[a\, dt -
(r^2+a^2) \,d\phi \right]^2
+ \frac{\cos^2\theta}{\Sigma} \left[b\, dt -
(r^2+b^2) \,d\psi \right]^2 \nonumber
\\ &&
+ \frac{1}{r^2 \Sigma } \left[ a  b \,dt - b
(r^2+a^2) \sin^2\theta\,d\phi
- a (r^2+b^2)
\cos^2\theta\,d\psi \right]^2,
\end{eqnarray}
where the metric functions are given by
\begin{eqnarray}
f& =&\frac{\left(r^2 + a^2\right)\left(r^2 +
b^2\right)}{r^2\Sigma} - \frac{2M\Sigma - Q^2}{\Sigma^2}\,,~~~~~~\Sigma =  r^2+ a^2 \cos^2\theta + b^2 \sin^2\theta\,,\nonumber \\[2mm]
\Delta &= &\left(r^2 + a^2\right)\left(r^2 +
b^2\right)+ 2 a b\, Q + Q^2 - 2 M r^2\,,
\label{gsugrametfunc}
\end{eqnarray}
the parameters $ M $ and $ Q $  are  related to the physical mass and electric charge of the black hole, whereas  $ a $ and $ b $ are two independent rotation parameters. The metric determinant is given by
\begin{equation}
\sqrt{-g}= r \Sigma \sin\theta\,\cos\theta\,.
\label{5ddeterminant}
\end{equation}
It is straightforward to check that this metric and the two-form field  $ F=dA $, where
\begin{equation}
A= -\frac{\sqrt{3}\,Q}{2 \Sigma}\,\left(dt - a
\sin^2\theta\,d\phi - b \cos^2\theta \,d\psi
\right)
\label{sugrapotform1}
\end{equation}
is the potential one-form of the electromagnetic field supporting the metric,  satisfy the equation of motions derived from  the action of  five-dimensional minimal  ungauged supergravity
\begin{eqnarray}
S&=& \int d^5x \sqrt{-g} \left(R -\frac{1}{4}\, F_{\alpha \beta}F^{\alpha \beta}
+\frac{1}{12\sqrt{3}}\,\epsilon^{\mu\nu \alpha\beta\lambda}F_{\mu\nu} F_{\alpha\beta}A_{\lambda}\right).
\label{5sugraaction}
\end{eqnarray}
The locations of  the outer and inner horizons of the black hole  are determined by the real  roots of the equation $ \Delta = 0 $. Thus, we find that
\begin{equation}
r_\pm^2 =\frac{1}{2}\left[-(a^2+b^2-2M)\pm \sqrt{(a^2+b^2-2M)^2-4(ab+Q)^2} \right],
\label{5horizons}
\end{equation}
where  $ r_{+}^2 $ corresponds to the radius of the outer (the event) horizon, while $ r_{-}^2 $  gives the radius of the inner Cauchy horizon. It follows that for the extremal horizon, $ r_{+}^2= r_{-}^2\,, $  there exist two simple relations  between the parameters of the black hole, which are given by
\begin{eqnarray}
M &= & \frac{(a+b)^2}{2} + Q\,,~~ or ~~M =\frac{(a-b)^2}{2}- Q\,.
\label{extreme2}
\end{eqnarray}

In the following we will also need the inverse components of  metric (\ref{gsugrabh}), which are given by
\begin{eqnarray}
g^{00}&=& -1 + \frac{(r^2+b^2)\left[Q^2- 2 M (r^2+a^2)\right]+  a^2 Q^2}{\Delta \Sigma}\,,~~~~~~ g^{11}=\frac{\Delta}{r^2\Sigma}\,,
\nonumber \\[3mm]
g^{22}&=&\frac{1}{\Sigma}\,,~~~~~~g^{33}=\frac{1}{\Sigma}\left[\frac{1}{\sin^2\theta} + \frac{(r^2+b^2)( b^2-a^2) -2b (a Q + b M)}{\Delta}\right],\nonumber\\[3mm]
g^{44}&=&\frac{1}{\Sigma}\left[\frac{1}{\cos^2\theta} + \frac{(r^2+a^2)( a^2-b^2)-2a (b Q+a M)}{\Delta}\right],~~~
g^{34}= -\frac{2 a  b M+ (a^2+b^2)\,Q}{\Delta \Sigma}\,,
\nonumber \\[3mm]
g^{03}&=&-\frac{(2 a M + b Q)(r^2 + b^2)- a Q^2}{\Delta \Sigma}\,,~~~~ g^{04} = -\frac{(2 b M + a Q)(r^2 + a^2)- b Q^2}{\Delta \Sigma}\,\,.
\label{contras}
\end{eqnarray}
It is easy to see that the stationary and bi-azimuthal isometries of this  metric are described  by three commuting Killing vectors
\begin{equation}
{\xi}_{(t)}= \frac{\partial}{\partial t}\,, ~~~~ {\xi}_{(\phi)}=
\frac{\partial}{\partial \phi} \, , ~~~~ { \xi}_{(\psi)}=
\frac{\partial}{\partial \psi }\,, \label{5dkilling}
\end{equation}
which can be used to define a family of locally nonrotating observers. The 5-velocity unit vector of these observers is given by
\begin{equation}
u^{\mu}= \alpha \left(
\xi_{(t)}^{\mu}+\Omega_{a}\xi_{(\phi)}^{\mu}
+\Omega_{b}\xi_{(\psi)}^{\mu}\right),
\end{equation}
where $\alpha $  is determined by the condition $ u^2=-1 $.  The defining relations $ u\cdot {\xi}_{(\phi)}=0 $ and $ u\cdot {\xi}_{(\psi)}=0 $  allow us  to determine the coordinate angular velocities $ \Omega_{a} $ and   $ \Omega_{b} $ of the observers (see e.g. \cite{aliev4} for some details). Performing straightforward calculations, we find that
\begin{eqnarray}
\label{5dangve2a}
 \Omega_{a} & =& \frac{\left(r^2+b^2\right) \left(2 a M + b Q\right) - a Q^2}
{\Delta  \Sigma + 2 M \left(r^2+a^2\right)\left(r^2+b^2 \right) - Q^2 \left(r^2+a^2+ b^2\right)}
 \,\,, \\ [5mm]
\Omega_{b} & =& \frac{\left(r^2 + a^2\right) \left(2 b M + a Q\right) - b Q^2}
{\Delta  \Sigma + 2 M \left(r^2+a^2\right)\left(r^2 + b^2 \right) - Q^2 \left(r^2+a^2+ b^2\right)}\,\,.
\label{5dangve2b}
\end{eqnarray}
At large distances, as follows from these expressions, the  bi-dragging property of the metric is governed by the remarkably simple formulas
\begin{eqnarray}
\Omega_{a} & =&  \frac{2 a M + b Q}{r^4}  + \mathcal{O}\left(\frac{1}{r^6}\right)\,\,, ~~~~~ \Omega_{b} = \frac{2 b M + a Q}{r^4}  + \mathcal{O}\left(\frac{1}{r^6}\right).
\label{5dangvel3}
\end{eqnarray}
We note that for vanishing rotation parameter $ a=0 $ (or $ b=0 $), the bi-dragging  still occurs due to the electric charge of the black hole. The effect disappears at spatial infinity, while it increases towards the horizon and for  $\Delta = 0 $,  expressions (\ref{5dangve2a}) and  (\ref{5dangve2b}) reduce to the angular velocities of the horizon \cite{cclp}.  We have
\begin{equation}
\Omega_{a(+)} = \frac{2 \pi^2}{\cal A}\cdot\frac{a(r_{+}^2 + b^2) + b Q}{r_{+}}
\,,~~~~~~~\Omega_{b(+)} = \frac{2 \pi^2}{\cal A}\cdot\frac{b(r_{+}^2 + a^2) + a Q}{r_{+}}\,,
\label{Hvelocity}
\end{equation}
where the horizon area $ {\cal A} $ is given by
\begin{eqnarray}
{\cal A} &=& \frac{2 \pi^2  \left[(r_{+}^2  + a^2)(r_{+}^2  + b^2) + a b Q\right]}{r_{+}}\,.
\label{area}
\end{eqnarray}
With these quantities in mind, we can now introduce a co-rotating Killing vector defined  as  follows
\begin{equation}
\chi = {\bf \xi}_{(t)}+ \Omega_{a(+)}\,{\bf \xi}_{(\phi)}+
\Omega_{b(+)} \,{\bf \xi}_{(\psi)}\,.
\label{5dkhorizon}
\end{equation}
It is straightforward to show that the norm of this vector vanishes on the horizon, showing that  it coincides with the null geodesic generators of the horizon. Using this vector, one can calculate the surface gravity $ \kappa $ of the horizon  and hence its Hawking temperature $ T_H $. We find that
\begin{eqnarray}
T_H &=& \frac{\kappa}{2 \pi} = \frac{\pi (r_{+}^2 - r_{-}^2)}{{\cal A}}\,,
\label{Htemp}
\end{eqnarray}
where we have used  expressions (\ref{5horizons}) and (\ref{Hvelocity}). The co-rotating Killing vector can  also be used to calculate the electrostatic potential of the horizon. Indeed, by means of potential one-form (\ref{sugrapotform1}) and expressions  (\ref{Hvelocity}), we find that the electrostatic potential of the horizon, relative to an infinitely  distant observer, is given by
\begin{eqnarray}
\Phi_H & = - & A \cdot \chi  = \frac{\sqrt{3}\, \pi^2 Q  r_+}{{\cal A}}\,.
\label{hpot}
\end{eqnarray}
Remarkably, the CCLP  metric in (\ref{gsugrabh}) admits hidden symmetries which are generated by a second-rank Killing tensor \cite{dkl, ad}, in addition to its global symmetries given by  Killing vectors (\ref{5dkilling}). As a consequence, the geodesic and  scalar field equations  separate  in this metric, ensuring their complete integrability. Below, we proceed with  the separation of variables in the Klein-Gordon equation for a charged massless scalar field.\\

\section{Klein-Gordon equation}

Let us consider a charged massless scalar field which obeys the Klein-Gordon equation $  D^{\mu}D_{\mu} \Phi= 0 $, where $ D_{\mu}= \nabla_{\mu}- ie A_{\mu} $, and $ \nabla_{\mu} $  is the covariant derivative operator with respect to metric  (\ref{gsugrabh}). Using expression (\ref{5ddeterminant}) and  the contravariant components of the metric given in (\ref{contras}),  it is straightforward to show  that this equation separates for the solution ansatz of the form
\begin{equation}
\Phi=e^{-i \omega t + i m_\phi  \phi +i m_\psi  \psi} S(\theta) R(r)\,,
\label{sansatz}
\end{equation}
where $ m_\phi $ and $ m_\psi $ are  ``magnetic" quantum numbers  associated with  rotation in the   $ \phi $ and $ \psi $ directions. The angular function $ S(\theta) $  obeys  the equation
\begin{eqnarray}
&&\frac{1}{\sin2\theta}\frac{d}{d\theta}\left(\sin2\theta\frac{d S}{d\theta}\right) +\left[\lambda -\omega^2 \left(a^2 \sin^2\theta + b^2 \cos^2\theta\right)
- \frac{m_\phi^2}{\sin^2\theta} -  \frac{m_\psi^2}{\cos^2\theta}
\right] S = 0\,,
\label{angular1}
\end{eqnarray}
where we have used  the freedom of shifting the separation constant, $\lambda \rightarrow \lambda + const $. As is known \cite{fs}, this equation when accompanied by regular boundary conditions at singular points $ \theta=0 $ and  $ \theta=\pi/2 $ defines a Sturm-Liouville problem. The associated  eigenvalues are $ \lambda=\lambda_{l}(\omega) $,  where $ l $ is an integer which can be thought of as an ``orbital" quantum number. The solution  is given
by the five-dimensional spheroidal functions $ S(\theta)= S_{\ell\, m_\phi m_\psi}(\theta|a \omega\,, b\omega) $, which form a complete set  over the integer $ l $.  For nonvanishing rotation parameters, but for  $  a^2 \omega^2\ll 1 $ and  $  b^2 \omega^2 \ll 1 $, one can show that
\begin{eqnarray}
\lambda &=& l(l+2) + \mathcal{O}\left(a^2 \omega^2,  b^2 \omega^2 \right),
\label{eigenv1}
\end{eqnarray}
where $ l $ must obey the condition $ l \geq m_\phi +m_\psi $ \cite{fs}.

The radial equation for $ R(r)$, by performing a few algebraic manipulations, can be cast in the form
\begin{equation}
\frac{\Delta}{r} \frac{d}{d r}\left(\frac{\Delta}{r} \frac{d R}{d r}\right) +U(r)\,R=0\,,
\label{radial1}
\end{equation}
where
\begin{eqnarray}
\label{rpotential}
&&U(r)=-\Delta \left[\lambda - 2\omega(a m_\phi +  b m_\psi) + \frac{\left(a b\, \omega-b m_\phi-a m_\psi\right)^2}{r^2}\right]
+\frac{\left[(r^2+a^2)(r^2+b^2)+ab\,Q \right]^2}{r^2} \times \nonumber
\\[4mm]
&&\left\{\omega- \frac{m_\phi \left[a(r^2+b^2)+b Q\right]}{(r^2+a^2)(r^2+b^2)+a b Q } - \frac{m_\psi \left[b(r^2+a^2)+a Q\right]}{(r^2+a^2)(r^2+b^2)+a b Q}
-\frac{\sqrt{3}}{2}\,\frac{ e Q r^2}{(r^2+a^2)(r^2+b^2)+a b Q }
\right\}^2.\nonumber \\
\label{radpot1}
\end{eqnarray}
For vanishing electric charge, $ Q=0 $, these expressions go over into those obtained in \cite{fs}. They also agree with the vanishing cosmological constant limit of the expressions given in \cite{ad}. Next, it  proves useful to transform the radial equation into a Schr\"{o}dinger  form. For this purpose, we introduce a new radial function $ \mathcal{R} $  and a new radial coordinate  $ r_* \,$, which are defined by the relations
\begin{eqnarray}
R &=& \left[\frac{r}{(r^2+a^2)(r^2+b^2)+a b Q}\right]^{1/2} \,\mathcal{R}\,,~~~~~~\frac{dr_*}{dr}=\frac{(r^2+a^2)(r^2+b^2)+a b Q}{\Delta}\,\,.
\label{newradfcord}
\end{eqnarray}
Using these definitions, we rewrite the radial equation (\ref{radial1}) in the   Schr\"{o}dinger form
\begin{eqnarray}
\frac{d^2 \mathcal{R}}{dr_*^2} +V(r) \mathcal{R}=0\,,
\label{radial2}
\end{eqnarray}
where the ``effective" potential  is given by
\begin{eqnarray}
&&V(r)= -\frac{\Delta  \left\{r^2 \left[\lambda - 2\omega(a m_\phi +  b m_\psi) \right]+\left(a b\,\omega-b m_\phi-a m_\psi\right)^2\right\}}{\left[(r^2+a^2)(r^2+b^2)+a b Q\right]^2}
-\frac{\Delta}{2 r Z^{3/2}}\,\frac{d}{d r}\left(\frac{\Delta}{r Z^{3/2}}\,\frac{d Z}{d r}\right) + \nonumber
\\[2mm] &&
\left\{\omega- \frac{m_\phi \left[a(r^2+b^2)+b\,Q\right]}{(r^2+a^2)(r^2+b^2)+a b Q } - \frac{m_\psi \left[b(r^2+a^2)+a Q\right]}{(r^2+a^2)(r^2+b^2)+a b Q}
-\frac{\sqrt{3}}{2}\,\frac{e Q r^2}{(r^2+a^2)(r^2+b^2)+a b Q }\right\}^2 \nonumber\\[2mm]
\label{effective}
\end{eqnarray}
and we have also used the notation
\begin{equation}
Z=\left[(r^2+a^2)(r^2+b^2)+a b Q\right]r^{-1}.
\end{equation}
We are now interested in the behavior of the radial equation in the asymptotic regions, at spatial infinity  $ r_* \rightarrow \infty $ and  at the horizon
$ r_* \rightarrow - \infty\, (r \rightarrow r_+ ),$ where the effective potential  (\ref{effective}) takes the form
\begin{equation}
V(r)\rightarrow \left\{
\begin{array}{ll}
\left(\omega-m_\phi \Omega_{a(+)}-m_\psi \Omega_{b(+)}-e \Phi_{H}\right)^2, & ~~~~~~~~~   r_* \rightarrow - \infty\\
\omega^2, & ~~~~~~~~~  r_* \rightarrow \infty
\end{array}
\right.
\label{asymppot}
\end{equation}
Meanwhile, in the intermediate region it acts as a barrier, resulting in scattering processes of radial waves. In general, these asymptotic relations  allow one to distinguish two classes of solutions to  the radial wave equation: (i) the first class of solutions  represents a wave originating at infinity (or being purely ingoing at the horizon), (ii) the second class of solutions corresponds to a wave originating in the past horizon (or being purely outgoing at infinity). The classical scattering process, which is the case under consideration, must be  represented by the first class of solutions. That is, we have the following  asymptotic behavior
\begin{equation}
\mathcal{R} \rightarrow \left\{
\begin{array}{ll}
T_A\,e^{-i \left(\omega-m_\phi \Omega_{a(+)}-m_\psi \Omega_{b(+)}-e \Phi_{H}\right) r_*}, & ~~~~~~~~~   r_* \rightarrow - \infty\\
e^{-i\omega r_*}+ R_A\,e^{i\omega r_*}, & ~~~~~~~~~  r_* \rightarrow \infty
\end{array}
\right.
\label{bouconds}
\end{equation}
where $ T_A $ and  $ R_A $  are the transmission and  reflection amplitudes respectively.  The complex-conjugate of these asymptotic forms  corresponds to  the associated complex-conjugate solution of  equation (\ref{radial2}) as  the effective potential $ V(r) $ is a real quantity. Clearly, these two solutions are linearly independent and using the constancy of their Wronskian, we find that   the transmission and  reflection amplitudes obey the  relation
\begin{equation}
| R_A |^2=1-\frac{\omega-\omega_p}{\omega}\,| T_A |^2,
\label{rtampls}
\end{equation}
where have introduced  the threshold frequency
\begin{equation}
\label{omegap}
\omega_p=m_\phi \Omega_{a(+)} + m_\psi \Omega_{b(+)} + e\Phi_{H}.
\label{porog}
\end{equation}
It follows  that for the frequency range given by the inequality
\begin{equation}
0<\omega < \omega_{p}\,,
\label{fbound}
\end{equation}
the reflected wave has greater amplitude than the incident one, $ | R_A |^2 > 1, $ i.e. the superradiant effect appears. We note that the presence of the electric charge changes the threshold frequency of superradiance. This occurs not only due to the nonvanishing electrostatic potential of the horizon but also because of  the gravimagnetic bi-dragging contribution to its angular velocities.

\section{Solutions}

The analysis of singularity structure of the radial equation (\ref{radial1}) reveals that solutions to this equation possess an essential singularity. This means that  one can not use the familiar techniques, employed in the theory of ordinary linear differential equations, to find the general solutions to this equation (see e.g. \cite{ince}). On the other hand, one can certainly find such solutions to some approximated  versions of this equation, which are applicable in various regions of the spacetime. In what follows, we are  interested in solutions  at low frequencies i.e., when the Compton wavelength of the scalar particle  is much larger than the horizon radius of the black hole. Following the work of Starobinsky \cite{starobinsky}, we divide  the spacetime into the near-horizon and far regions  and approximate  equation (\ref{radial1}) for each of these regions. Solving then the resulting  equations with appropriate boundary conditions, we assume that the orbital quantum number $ l $  is nearly integer, thereby avoiding the appearance of solutions with logarithmic terms.
This allows us to perform the matching  of the solutions under consideration  in the overlap between the near and far regions and thus obtaining  the complete solution at  low frequencies. Below, we discuss these equations and solutions to them  as well as the matching procedure in the overlap region.

\subsection{Near-Region}

In the region near the horizon, $ r- r_+ \ll 1/\omega $, and for low-frequency perturbations  $  r_+  \ll 1/\omega $, equation  (\ref{radial1}) takes the form
\begin{equation}
\frac{\Delta}{r} \frac{d}{d r}\left(\frac{\Delta}{r} \frac{d R}{d r}\right) +
\left(\frac{\omega- \omega_p}{2\pi^2}\,{\cal A}\right)^2 R -l(l+2) \Delta R=0\,,
\label{nearrad1}
\end{equation}
where we have  used  relations (\ref{area}) and (\ref{eigenv1}), assuming slow rotation as well. For future purposes,  we will henceforth assume that $ l $ is nearly integer, thus keeping in mind small corrections in (\ref{eigenv1}) and (\ref{rpotential}). Next, using a new dimensionless variable
\begin{equation}
z=\frac{r^2-r_+^2}{r^2-r_-^2}\,,
\end{equation}
one can show that equation (\ref{nearrad1}) reduces to the hypergeometric type equation
\begin{equation}
z (1-z) \frac{d^2R}{dz^2} +(1-z)\frac{d R}{dz} +\left[ \frac{1-z}{z}\,\Omega^2 -\frac{\ell(\ell+2)}{4(1-z)}\right]R=0\,,
\label{nearrad2}
\end{equation}
where
\begin{equation}
\Omega= \frac{\omega-\omega_p}{4 \pi T_H}\,.
\label{newsuperf}
\end{equation}
This equation can be solved in a standard way by the ansatz
\begin{equation}
R(z)=z^{i \Omega}\,(1-z)^{1 + l/2}\,F(z)\,,
\end{equation}
where  $ F(z)= F(\alpha\,, \beta\,,\gamma, z) $ is the hypergeometric  function, obeying the equation
\begin{eqnarray}
&& z(1-z)\frac{d^2 F}{dz^2} +\left[\gamma-(\alpha +\beta+1)\,z\right] \frac{d F}{dz} - \alpha \beta F = 0\,,
\label{hyperg1}
\end{eqnarray}
and the parameters are given by
\begin{equation}
\alpha= 1+l/2 + 2 i \Omega\\,,~~~~~~~\beta= 1+ l/2\,,~~~~~~~
\gamma=1+2i\Omega\,.
\label{nearradpara}
\end{equation}
Thus, the general solution to equation (\ref{nearrad2}) can be written in terms of two linearly independent solutions of equation (\ref{hyperg1}). We need the physical solution that reduces to the ingoing  wave at the horizon, $ z\rightarrow 0 $. It is given by
\begin{equation}
R(z)= A_{(+)}^{in} \,z^{-i \Omega}\,(1-z)^{1+ l/2}\,F\left(1+ l/2\,, 1+ l/2 - 2 i \Omega\,, 1-2i\Omega\,, z\right)\,,
\label{nearphys}
\end{equation}
where $ A_{(+)}^{in} $ is a constant. Furthermore, in an overlapping region  the  large $ r $   behavior of  this solution should be compared with the small  $ r $ behavior of the far-region solution.  Therefore, we also need the large $ r $  $  (z \rightarrow 1) $  limit of solution (\ref{nearphys}) which can be easily found by using the pertinent modular properties of the hypergeometric functions \cite{abramowitz}. We find that the large $ r $ behavior of the near-horizon region solution is given by
\begin{eqnarray}
R \sim   A_{(+)}^{in} \,\Gamma(1-2i\Omega)\left[\frac{\Gamma(-l - 1)\,(r_+^2- r_-^2)^{1+ l/2}}{\Gamma(- l/2)\,\Gamma(- l/2-2i\Omega)}\,\,r^{-2 - l} + \frac{\Gamma(l+1)\,(r_+^2- r_-^2)^{-l/2}}{\Gamma(1+ l/2)\,\Gamma(1+ l/2-2i\Omega)}\, \,r^{l}\right].\nonumber \\
\label{larnear}
\end{eqnarray}
It is important to note that in this expansion the first term inside the square bracket requires a special care  for $ l $ approaching the integer values as the quotient of two gamma functions $ \Gamma(-l-1)/\Gamma(- l/2) $ becomes divergent for some values of $ l $. We will return to this issue  in more detail below.

\subsection{Far-Region}

In the far-region,  $r-r_+\gg r_+ $, equation (\ref{radial1}) can be approximated by
\begin{equation}
\frac{d^2R}{dr^2}+\frac{3}{r} \frac{dR}{dr}+\left[\omega^2-\frac{\ell(\ell+2)}{r^2} \right]R=0\,.
\label{radfar1}
\end{equation}
Here $ l $ is again supposed to be nearly, but not exactly, integer by taking into account small corrections in the region under consideration, including  the Newtonian term  $ \sim  \omega^2 r_+^2 /r^2  $ in five dimensions.

Using the ansatz  $ R = u/r $  and rescaling  the radial variable as $ x= \omega r, $ one can show that equation (\ref{radfar1}) reduces to the standard  Bessel equation given by
\begin{eqnarray}
&& x^2  \frac{d^2 u}{dx^2} + x  \frac{d u}{dx} - \left[x^2 -(l+1)^2\right] u  = 0\,.
\label{besseleq}
\end{eqnarray}
As is known \cite{abramowitz}, the general solution of this equation
is  a linear combination of the Bessel and  Neumann functions. We have
\begin{eqnarray}
R(r)=\frac{1}{r}\left[A_{\infty}\, J_{l+1}(\omega r)+  B_{\infty}\, N_{l+1} (\omega r) \right],
\label{bessel1}
\end{eqnarray}
where $ A_{\infty} $ and  $ B_{\infty} $ are constants. Though this solution refers only to large $ r $  region, but for small $ x $  ($\omega r \ll 1$) it might also have a limiting behavior, which indicates on an overlapping regime of validity with the  large $ r $  form of the near-horizon  solution (\ref{larnear}). For  small $  \omega r $, using  the asymptotic forms of the  Bessel and  Neumann functions,  we find that
\begin{eqnarray}
R(r) \sim  A_{\infty}\,\left(\frac{\omega}{2}\right)^{l+1}
\frac{r^l}{\Gamma(l+2)} -  B_{\infty}  \left(\frac{2}{\omega}\right)^{l+1 }\,\frac{ \Gamma(l+1)}{\pi}\,\, r^{-2 - l}\,.
\label{asypmbess1}
\end{eqnarray}
For some further purposes, it may also be  useful to know  the large  $  \omega r $ behavior of solution (\ref{bessel1}), which is given by
\begin{eqnarray}
R(r) \sim \frac{1}{\sqrt{2\pi \omega r^3}}\left[\left(A_{\infty} + i  B_{\infty}\right)e^{\frac{i \pi}{2} (l+ \frac{3}{2})} e^{-i\omega r} +    \left(A_{\infty} - i  B_{\infty}\right) e^{- \frac{i \pi}{2} (l+ \frac{3}{2})} e^{i\omega r}\right],
\label{asypmbess2}
\end{eqnarray}
where, as expected, the first term refers to an ingoing wave and  the second term corresponds to an outgoing wave.

\subsection{Matching Procedure}

With the above discussion of solutions, referring to the near-horizon and far regions of the spacetime, it becomes clear that  the construction of the complete low-frequency solution  for the radial waves  requires  a matching procedure in an intermediate region. Before doing this several comments are in order. Since the gamma function develops the pole structure  when its argument is a negative integer, it is easy to see that the quotient of gamma functions $ \Gamma(-l-1)/\Gamma(- l/2) $ appearing in expression (\ref{larnear}) diverges  for odd integer values of $ l $. Consequently, solutions with logarithmic terms will inevitably appear. This makes the matching procedure impossible for odd $ l $, as noted in \cite{hwang}.  However, assuming that  $ l $ is not exactly, but nearly integer one can avoid the appearance of solutions with logarithmic terms and proceed with the  matching procedure. This is the reason why we introduce the ``nearly integer" $ l $  in the above description of the solutions (see also \cite{page}).

With this in mind, we  compare equations  (\ref{larnear}) and (\ref{asypmbess1}) and see that there  exists an overlapping regime of validity  ($ r_+ \ll r-r_+\ll 1/\omega $) for the near-horizon and far region solutions. Performing  the matching in this regime,  we find that the defining  amplitude ratios are given by
\begin{eqnarray}
\label{aa}
\frac{A_{(+)}^{in}}{A_{\infty}} &= &\left(\frac{\omega}{2}\right)^{l+ 1}\frac{(r_+^2 - r_-^2)^{l/2}\, \Gamma(1+l/2)}{\Gamma(l+1)\Gamma(l+2)}\,
\frac{\Gamma(1+ l/2 -2i \Omega)}{\Gamma(1- 2i \Omega)}\,,\\[5mm]
\frac{B_{\infty}}{A_{\infty}} &= & - \pi \,\left(\frac{\omega}{2}\right)^{2(l+ 1)} \frac{(r_+^2- r_-^2)^{l+1}\,\Gamma(1+l/2)}{\Gamma^2(l+1)\Gamma(l+2)}\, \frac{\Gamma(-l-1)}{\Gamma(- l/2)} \, \frac{\,\Gamma(1+l/2-2 i \Omega)}{\Gamma(- l/2 -2i \Omega)}\,.
\label{ba}
\end{eqnarray}
We are now in position to proceed  with the superradiant  instability of the rotating black hole by placing  a reflecting mirror around it.

\section{Reflecting Mirror and Negative Damping}

As we have  described  in the introduction, one of the most striking application of the superradiant effect in four dimensions  amounts to exploring the black hole-mirror system, which under certain condition acts as a black hole bomb \cite{press1}. In this section,  we wish to explore this phenomenon in five dimensions, using the model which consists of a rotating black hole of minimal  ungauged supergravity  \cite{cclp} and a reflecting mirror located at a large  distance $ L $ from the black hole $ (L\gg r_+)$ .
We assume that the mirror perfectly reflects low-frequency scalar waves, so that on the surface of the mirror one must impose the vanishing  field condition. This, by equation (\ref{bessel1}), yields
\begin{eqnarray}
A_{\infty}\, J_{l+1}(\omega L)+  B_{\infty}\, N_{l+1}(\omega L)=0\,.
\label{vanishf}
\end{eqnarray}
This condition, when combined  with that requiring a purely ingoing wave at the horizon, defines a characteristic-value problem for the confined  spectrum of the low-frequency solution, discussed above. Such a spectrum would be  quasinormal with complex frequencies whose imaginary part describes the damping of modes,  as can be seen from equation (\ref{sansatz}). When the imaginary part is positive, a characteristic mode undergoes  exponential growth ({\it the negative damping}). In this case, the system will develop instability, creating  a black hole bomb.

Comparing now equations  (\ref{ba}) and (\ref{vanishf}), we obtain the defining transcendental equation for the frequency spectrum
\begin{eqnarray}
\frac{J_{l+1}(\omega L)}{N_{l+1}(\omega L)} = \pi \,\left(\frac{\omega}{2}\right)^{2(l+ 1)} \frac{(r_+^2- r_-^2)^{l+1}\,\Gamma(1+l/2)}{\Gamma^2(l+1)\Gamma(l+2)}\, \frac{\Gamma(-l-1)}{\Gamma(- l/2)} \, \frac{\,\Gamma(1+l/2-2 i \Omega)}{\Gamma(- l/2 -2i \Omega)}\,,
\label{defeq}
\end{eqnarray}
which can be solved by iteration in the low-frequency approximation. Let us assume that the solution to this equation can be written in the form
\begin{equation}
\omega= \omega_n + i \delta\,,
\label{complexf}
\end{equation}
where  $n $  is a non-negative integer,  $\omega_n $ describes  the discrete frequency spectrum of free modes and  $ \delta $ is supposed to be a small damping parameter, representing a ``response" to the ingoing wave condition at the horizon. Using this  in equation (\ref {defeq}) it is easy to see that, in lowest approximation,  $\omega_n $ is simply given by the real roots of the Bessel function. Thus, we have
\begin{eqnarray}
\omega_n = \frac{j_{{l+1}\,,\,n}}{L}\,,
\label{fundmod}
\end{eqnarray}
where the quantity  $j_{{l+1}\,,\,n} $  represents the $n$-th root (greater than zero) of the equation $J_{l+1}(\omega_n L)=0 $. A detailed list of  these roots can be found  in \cite{abramowitz}. They can also be easily tabulated using Mathematica. On the other hand, for large overtones of the fundamental frequency $ (n \gg 1) $ one can appeal to the asymptotic form of the Bessel function, which gives the simple formula
\begin{eqnarray}
j_{{l+1}\,,\,n} \simeq \pi \left(n + l/2\right)\,.
\label{asymfundmod}
\end{eqnarray}
It should be noted that  formula (\ref{fundmod}) generalizes to five dimensions the familiar flat spacetime result  for the frequency spectrum in an infinitely deep spherical potential well \cite{LL}.

Next, substituting  equations (\ref{complexf}) and (\ref{fundmod}) in equation (\ref{defeq}) and performing a few algebraic manipulations, to first order  in $ \delta $, we find that the damping parameter is given by
\begin{eqnarray}
\delta &=&  -\frac{i \pi}{L}\, \,\frac{N_{l+1}(j_{{l+1}\,,\,n})}{J^{\prime}_{l+1}(j_{{l+1}\,,\,n})}\,
\left( \frac{j_{{l+1}\,,\,n}}{2L}\right)^{2(l+1)}\,
\frac{(r_+^2- r_-^2)^{l+1}\, \Gamma(1+l/2)}{\Gamma^2(l+1)\,\Gamma(l+2)}\times
\nonumber
\\[3mm] &&
\frac{\Gamma(-l-1)}{\Gamma(- l/2)} \,
\frac{\,\Gamma(1+l/2-2 i \Omega)}{\Gamma(- l/2 -2i \Omega)}\,.
\label{sigma}
\end{eqnarray}
Here the prime denotes the derivative of the Bessel function with respect to its argument and the quantity $ \Omega $, as follows from equation (\ref{newsuperf}), is given by
\begin{equation}
\Omega=\frac{r_+^{3}}{2}\, \frac{\omega_n-\omega_p}{r_+^2-r_-^2}\,.
\label{superff}
\end{equation}
Comparing this expression  with that given in (\ref{fundmod}), we see that
the superradiant effect crucially depends on the distance $ L $ at which the mirror is placed just as in four dimensions \cite{cardoso1}. That is, for a critical distance governing the fundamental frequency, the effect ceases to exist. To proceed further, it is useful to simplify separately  the product of the quotients of  gamma functions in the second line of equation (\ref{sigma}).  Using the well known relation $ \Gamma(z)\Gamma(1-z)=\pi/\sin{\pi z} $,  it is straightforward to show that
\begin{eqnarray}
\label{simpli1}
\frac{\Gamma(-l-1)}{\Gamma(- l/2)} &=& -  \frac{1}{2 \cos(\pi l/2)}\,\frac{\Gamma (1+l/2)}{\Gamma(l+2)}\,,\\[4mm]
\frac{\Gamma(1+l/2-2 i \Omega)}{\Gamma(- l/2 -2i \Omega)}\
& =& - \frac{1}{\pi}\, \left|\Gamma(1+l/2-2 i \Omega)\right|^2 \left[\sin(\pi l/2) \cosh(2\pi \Omega)
 \right. \nonumber \\[3mm]  & & \left.
 +\, i \cos(\pi l/2) \sinh(2\pi \Omega)\right].
\label{simpli2}
\end{eqnarray}
Substituting now these relations in equation (\ref{sigma}), we have
\begin{eqnarray}
\delta &=&  \frac{i}{2 L}\,\,\left|\frac{N_{l+1}(j_{{l+1}\,,\,n})}{J^{\prime}_{l+1}(j_{{l+1}\,,\,n})}\right|\,
\left( \frac{j_{{l+1}\,,\,n}}{2L}\right)^{2(l+1)}\,
\frac{(r_+^2 - r_-^2)^{l+1} \Gamma^2(1+l/2)}{\Gamma^2(l+1)\,\Gamma^2(l+2)}\times
\nonumber
\\[3mm] &&
\frac{\left|\Gamma(1+l/2-2 i \Omega)\right|^2}{\cos(\pi l/2)}  \left[\sin(\pi l/2) \cosh(2\pi \Omega) + i \cos(\pi l/2) \sinh(2\pi \Omega)\right],
\label{sigmaf}
\end{eqnarray}
where we have changed the  overall sign, taking the absolute value of  the quotient $\,\frac{N_{l+1}(j_{{l+1}\,,\,n})}{J^{\prime}_{l+1}(j_{{l+1}\,,\,n})}\,, $  since  it is always negative in the physically acceptable frequency range.
Recalling that here $ l $ is nearly integer, we can further simplify this equation  by specifying  $ l $. Let us now assume that $ l $  approaches either even or odd integers. That is, we consider the following cases;

(i)  $ l/2 = p + \epsilon $, where  $ p $ is a non-negative integer and $ \epsilon \rightarrow 0 $.  Substituting this in expression (\ref{sigmaf}), we find that its imaginary part vanishes  in the limit $ \epsilon\rightarrow 0 $, whereas the real part  is given by
\begin{eqnarray}
\delta &=& - \pi \Omega \,\left|\frac{N_{2p+1}(j_{{2p+1}\,,\,n})}{J^{\prime}_{2p+1}(j_{{2p+1}\,,\,n})}\right|
\left( \frac{j_{{2p+1}\,,\,n}}{2L}\right)^{2(2p+1)}\,
\frac{(r_+^2 - r_-^2)^{2p+1}}{L}  \left(  \frac{p!}{(2p)! (2p+1)!} \right)^2 \prod_{k=1}^{p}\left(k^2+4 \Omega^2 \right).\nonumber\\
\label{sigmaeven}
\end{eqnarray}
In obtaining this expression we have  used  the identity
\begin{eqnarray}
\left|\Gamma(1+l/2-2 i \Omega)\right|^2 = \frac{2\pi\Omega}{\sinh(2\pi \Omega)}\,\prod_{k=1}^{p}\left(k^2+4 \Omega^2 \right),
\label{id1}
\end{eqnarray}
which can be easily obtained from the pertinent  properties of gamma functions   \cite{abramowitz}. It should be noted that indeed in the case under consideration, there are no divergencies in expression (\ref{sigmaf}) when $ \epsilon \rightarrow 0 $, so that  throughout the calculations  one can simply set $ \epsilon $ equal to zero. Turning back to equation (\ref{sigmaeven}), we see that its sign is entirely determined by  the sign of the quantity $\Omega $,  becoming positive in the superradiant regime, $\Omega < 0 $. Thus,  for all modes of even $ l $ we have the negative damping effect, resulting in exponential growth of their amplitudes.

(ii)  $ l/2 = (p + 1/2) + \epsilon $, again  $ p $ is a non-negative integer and $ \epsilon \rightarrow 0 $. Inserting this in expression (\ref{sigmaf}), we need to consider the limit as $ \epsilon \rightarrow 0 $. After performing a few straightforward calculations, we obtain that
\begin{eqnarray}
\delta &=& - \left|\frac{N_{2p+2}(j_{{2p+2}\,,\,n})}{J^{\prime}_{2p+2}(j_{{2p+2}\,,\,n})}\right|\,
 \left(\frac{j_{{2p+2}\,,\,n}}{2L}\right)^{2(2p+2)}
\frac{(r_+^2 - r_-^2)^{2p+2}}{2L}  \nonumber
\frac{\Gamma^2(p+3/2)}{\Gamma^2(2p+2)\Gamma^2(2p+3)}\times
\\[3mm] &&
\left|\Gamma(p+3/2-2 i \Omega)\right|^2\left(\sinh(2\pi \Omega) + \frac{i}{\epsilon} \, \frac{\cosh(2\pi \Omega)}{\pi} \right).
\label{sigmareim}
\end{eqnarray}
Using the properties of gamma functions \cite{abramowitz}, resulting in the relations
\begin{eqnarray}
\Gamma\left(p+\frac{1}{2}\right) &= & \frac{\pi^{1/2}\,(2p)!}{2^{2p}\, p!}\,,\\
\left|\Gamma(p+3/2 - 2 i \Omega)\right|^2 & =& \frac{\pi}{\cosh(2\pi \Omega)}\,\prod_{k=1}^{p+1}\left[(k-1/2)^2+4 \Omega^2 \right],
\label{funct}
\end{eqnarray}
one can further simplify the  combination of gamma functions appearing in equation (\ref{sigmareim}). Finally, we have
\begin{eqnarray}
\delta &=& - \pi \left|\frac{N_{2p+2}(j_{{2p+2}\,,\,n})}{J^{\prime}_{2p+2}(j_{{2p+2}\,,\,n})}\right|\,
 \left(\frac{j_{{2p+2}\,,\,n}}{4L}\right)^{2(2p+2)}
\frac{(r_+^2 - r_-^2)^{2p+2}}{2L} \times
 \nonumber \\[3mm] &&
\frac{\left(\pi \tanh(2\pi \Omega) + i/\epsilon\,\right)}{\left[(p+1)! (2p+1)!\right]^{2}}
 \,\prod_{k=1}^{p+1}\left[(k-1/2)^2+4 \Omega^2 \right].
\label{sigmafinal}
\end{eqnarray}
It is easy see that this expression possesses  two important features: first, its real part  that describes the damping of the modes changes the sign in the superradiant regime,  $\Omega < 0 $. This means that all modes of odd $ l $ may become supperradiant as well, resulting in the instability of the system. Meanwhile, the sign changing  does not occur for the imaginary part, which is not sensitive to superradiance at all. Second, the imaginary part  involves $ 1/\epsilon\, $ type divergence as $ \epsilon \rightarrow 0 $. However, this divergence  can somewhat be smoothed out by using the fact that the quantity $ r_+ $ is indeed  small, in accordance with the regime of validity of  the low-frequency solution constructed above. Thus,  for a given radius $ L $ of the mirror and for the lowest mode $(p=0) $, the ratio $ (r_+^2 - r_-^2)^2/ \epsilon\, $ appearing in  the imaginary part can be fixed as finite, to  high accuracy. The accuracy considerably increases for higher modes, as can be seen from (\ref{sigmafinal}). This would result in a  small frequency-shift in the  spectrum. These arguments is further supported by a numerical analysis of expression (\ref{sigmafinal}).

In Table I we present the numerical results  for  a charged nonrotating black hole. For the extreme charge of the black hole, we have $ Q_e = r_+^2 \,$, as follows from expressions (\ref{5horizons}) and  (\ref{extreme2}), and we take $ L=1 $, for certainty. The calculations  are performed for the parameters $ r_+=0.01 $,~ $ e=10 \,$ and for the lowest modes  as  $ l $ approaches  even or odd integers.  We see that  the superradiant instability appears in both cases, when  the charge of the black hole is close to the extreme value. Meanwhile, for $\epsilon\rightarrow 10^{-7} $, the imaginary part of the damping parameter (in the odd $l$  case) represents a small frequency-shift in the spectrum.
Table II gives a summary of the numerical analysis of the damping parameter  for a singly rotating black hole with zero  electric charge, $Q=0 $. It follows that the superradiant  instability  occurs  to all $ l $ modes of scalar perturbations under consideration. Again,  we a have a small frequency-shift for the $ l=1$  mode, by  choosing $\epsilon\rightarrow 10^{-7} $.

\begin{table}
\label{5Dsugrastatic}
\caption{The damping parameter of quasinormal modes ($ p=0,\, n=1 $); the scalar field charge $e=10$,\, the black hole parameters  $a=b=0,\, r_+=0.01$ and $ q= Q/Q_e\,.$}
\begin{ruledtabular}
\begin{tabular}{l l l }
$q $ &$\delta_{\ell/2=p+\epsilon}\,,~~\epsilon\rightarrow 0 $  & $\delta_{\ell/2= (p+1/2)+\epsilon}\,,~~\epsilon\rightarrow 10^{-7} $ \\
\hline
0.1 & $-5.653\times 10^{-5}$  & $-2.992\times 10^{-8} -0.462 \times 10^{-7}$\,i/$\epsilon $\\
0.3 & $-4.035 \times 10^{-5}$ & $-2.248\times 10^{-8} -0.422 \times 10^{-7}$\,i/$\epsilon $    \\
0.5 & $-2.274\times 10^{-5}$  & $-1.408\times 10^{-8} -0.346 \times 10^{-7}$\,i/$\epsilon $\\
0.7 & $-3.911 \times 10^{-6}$  & $-5.802\times 10^{-9} -0.234 \times 10^{-7}$\,i/$\epsilon $\\
0.8 & $ +5.956 \times 10^{-6}$ & $-2.213\times 10^{-9} -0.165 \times 10^{-7}$\,i/$\epsilon $  \\
0.9 & $+16.221\times 10^{-6}$  & $+ 4.965\times 10^{-10} -0.087 \times 10^{-7}$\,i/$\epsilon $  \\
\end{tabular}
\end{ruledtabular}
\end{table}
\begin{table}
\label{5Dsugrarotating}
\caption{The damping parameter of quasinormal modes with $ m_\phi=1 $
($ p=1,\, n=1 $ in the even $l$ case,  while  $ p=0,\, n=1 $ in the odd $l$ case); the  black hole parameters  $r_+=0.01$,\, $ \alpha= a/r_+\,,$ $\,b=0\,$ and $ Q =0.$}
\begin{ruledtabular}
\begin{tabular}{l l l }
$\alpha $ &$\delta_{\ell/2=p+\epsilon}\,,~~\epsilon\rightarrow 0 $  & $\delta_{\ell/2= (p+1/2)+\epsilon}\,,~~\epsilon\rightarrow 10^{-7} $ \\
\hline
0.1 & $4.673\times 10^{-13}$  & $5.513\times 10^{-9} -0.115 \times 10^{-7}$\,i/$\epsilon $\\
0.2 & $1.788 \times 10^{-12}$ & $1.708\times 10^{-8} -0.124 \times 10^{-7}$\,i/$\epsilon $    \\
0.3 & $3.215\times 10^{-12}$  & $ 2.935\times 10^{-8} -0.143 \times 10^{-7}$\,i/$\epsilon $\\
0.33 & $3.675 \times 10^{-12}$  & $3.323\times 10^{-8} -0.150\times 10^{-7}$\,i/$\epsilon $\\
\end{tabular}
\end{ruledtabular}
\end{table}

Thus, we conclude that in the black hole-mirror model under consideration,
all $ l $ modes  of scalar perturbations  become unstable in the regime of superradiance, exponentially growing their amplitudes with characteristic time scale $ \tau=1/\delta $. In addition, the modes of odd $ l $ undergo small frequency-shifts in the spectrum.

\section{Conclusion}

The superradiant instabilities of black hole-mirror systems  as well as  small AdS black holes in four-dimensional spacetimes have been extensively  studied in \cite{cardoso1, cardoso2} by employing both analytical and numerical approaches. The analytical approach is  based on a matching procedure, first introduced by Starobinsky \cite{starobinsky}, that allows one to find the complete  low-frequency solution to the Klein-Gordon equation by matching  the near-horizon and far regions solutions in their overlap region. In our earlier work \cite{ad}, using a similar analytical approach we gave a quantitative description of the superradiant instability of   small rotating charged AdS black holes in five dimensions. In a recent development \cite{hwang}, this investigation was  continued for small Reissner-Nordstr\"{o}m-AdS  black holes in all spacetime dimensions. Here it was also pointed out that in  odd spacetime dimensions, the matching  procedure  used in \cite{ad} fails for some values of the orbital quantum number $l $,  thus making the use of numerical methods inevitable.

The purpose of this paper was to embark on a further investigation of the superradiant instability in five dimensions,  elaborating on the black hole bomb model which consists of a rotating black hole of five-dimenensional minimal ungauged supergravity \cite{cclp} and a reflecting mirror around it. In spite of some subtleties with the matching procedure in five dimensions, we have shown that  one can still successfully use the analytical approach to  give the quantitative  description of the black hole bomb model under consideration.

Our results  can be summarized as follows: After   demonstrating the full separability of the Klein-Gordon equation, we have discussed  the behavior of the radial wave equation in the asymptotic regions and derived the threshold inequality for superradiance. Next, focusing on low-frequency  perturbations and slow rotation, we have approximated the radial wave equation in the near-horizon and far regions of the spacetime and solved the resulting equations with appropriate boundary conditions in each of these regions separately. To avoid  the appearance of solutions with logarithmic terms, which do not comply with the matching procedure, we have assumed that the orbital quantum number $ l $ is not exactly, but nearly integer. With this in mind, we have  performed the  matching of the near-horizon and far regions solutions  in an intermediate region, thereby constructing the complete low-frequency solution to the Klein-Gordon equation.

In the black hole-mirror system, we have defined a characteristic-value problem for the confined (quasinormal) spectrum of the low-frequency solution and calculated the complex frequencies of the spectrum. We have found the general expression for the imaginary part (for the small damping parameter) of the quasinormal spectrum, which appeared to be a complex quantity. Next, taking the limit as $ l $  approaches  an even  integer, we have shown that  the imaginary part of the damping parameter vanishes identically, whereas its real part becomes positive in the superradiant regime. Thus,  all modes of even $ l $ undergo negative damping, resulting in exponential growth of their amplitudes. Meanwhile, in the limit as  $ l $  approaches  an odd  integer, the damping parameter remains complex whose real part is positive in the superradiant regime, thereby showing that   all modes of odd $ l $ become unstable as well. As for the imaginary part, its sign appears to be  not sensitive to superradiance at all. We have argued that to high accuracy, the imaginary part of the damping parameter can be considered as representing a small frequency-shift in the spectrum, as discussed at the end of Sec. V.

Finally, we have concluded  that  that in the five-dimensional black hole-mirror  system, all $ l $ modes  of scalar perturbations undergo  negative damping in the regime of superradiance, exponentially growing their amplitudes and thus creating the  black hole bomb effect in five dimensions.

\section{Acknowledgments}

The author thanks  Ekrem  \c{C}alk{\i}l{\i}\c{c} and H. H\"{u}sn\"{u} G\"{u}nd\"{u}z for their stimulating encouragement. He also thanks  Mengjie Wang for useful  correspondence. This work is supported by the Scientific and Technological Research Council of Turkey (T{\"U}B\.{I}TAK)  under the Research Project No. 110T312.

\end{document}